# A Reference Based, Tree Structured Time Synchronization Approach and its Analysis in WSN

Surendra Rahamatkar[1] and Dr. Ajay Agarwal[2]

[1]Dept. of Computer Sc., Nagpur Institute of Technology, Nagpur, India
`rahamatkar_s@rediffmail.com`,
[2]Dept. of MCA, Krishna Inst. of Engg. & Technology, Ghaziabad, India
`ajay.aagar@gmail.com`

## ABSTRACT

*Time synchronization for wireless sensor networks (WSNs) has been studied in recent years as a fundamental and significant research issue. Many applications based on these WSNs assume local clocks at each sensor node that need to be synchronized to a common notion of time. Time synchronization in a WSN is critical for accurate time stamping of events and fine-tuned coordination among the sensor nodes to reduce power consumption. This paper proposes a bidirectional, reference based, tree structured time synchronization service for WSNs along with network evaluation phase. This offers a push mechanism for (i) accurate and (ii) low overhead for global time synchronization. Analysis study of proposed approach shows that it is lightweight as the number of required broadcasting messages is constant in one broadcasting domain.*

## KEYWORDS

*Ad Hoc Tree Structure; Clock synchronization; Wireless sensor networks, Hierarchical sensor network.*

## 1. INTRODUCTION

Wireless sensor networks (WSNs) can be applied to a wide range of applications in domains as diverse as medical, industrial, military, environmental, scientific, and home networks [5]. Since the sensors in a WSN operate independently, their local clocks may not be synchronized with one another. This can cause difficulties when trying to integrate and interpret information sensed at different nodes. For instance, if a moving car is detected at two different times along a road, before we can even tell in what direction the car is going, the detection times have to be compared meaningfully. In addition, we must be able to transform the two time readings into a common frame of reference before estimating the speed of the vehicle. Estimating time differences across nodes accurately is also important in node localization. For example, many localization algorithms use ranging technologies to estimate inter-nodes distances; in these technologies, synchronization is needed for time-of-flight measurements that are then transformed into distances by multiplying with the medium propagation speed for the type of signal used such as radio frequency or ultrasonic. There are additional examples where cooperative sensing requires the nodes involved to agree on a common time frame such as configuring a beam-forming array and setting a TDMA (Time Division Multiple Access) radio schedule [6]. These situations mandate the necessity of one common notion of time in WSNs. Therefore, currently there is a huge research interest towards developing efficient clock synchronization protocols to provide a common notion of time.

Time synchronization of WSNs is crucial to maintain data consistency, coordination, and perform other fundamental operations. Further, synchronization is considered a critical problem for wireless ad hoc networks due to its de-centralized nature and the timing uncertainties introduced by the imperfections in hardware oscillators and message de-lays in both physical and medium access control (MAC) layers. All these uncertainties cause the local clocks of different nodes to drift away from each other over the course of a time interval.

The primary functionality of wireless sensor networks is to sense the environment and transmit the acquired information to base stations for further processing with secure time information [3]. The clock synchronization problem has been studied thoroughly in the areas of Internet and local area networks (LANs) for the last several decades. Many existing synchronization algorithms rely on the clock information from Global Positioning System (GPS). However, GPS-based clock acquisition schemes exhibit some weaknesses: GPS is not ubiquitously available and requires a relatively high-power receiver, which is not possible in tiny and cheap sensor nodes. This is the motivation for developing software-based approaches to achieve in-network time synchronization. Among many protocols that have been devised for maintaining synchronization, Network Time Protocol (NTP) [7] is outstanding owing to its ubiquitous deployment, scalability, robustness related to failures, and self-configuration in large multi-hop networks. Moreover, the combination of NTP and GPS has shown that it is able to achieve high accuracy on the order of a few microseconds [8]. However, NTP is not suitable for a wireless sensor environment, since WSNs pose numerous challenges of their own; to name a few, limited energy and bandwidth, limited hardware, latency, and unstable network conditions caused by mobility of sensors, dynamic topology, and multi-hopping. The most of the time synchronization protocols differ broadly in terms of their computational requirements, energy consumption, precision of synchronization results, and communication requirements [1].

In the paper, we propose a more effective, lightweight multi-hop tree structured referencing time synchronization (TSRT) approach with the goal of achieving a long-term network-wide synchronization with minimal Message Exchanges and exhibits a number of attractive features such as highly scalable and lightweight.

The whole paper is organized in six Sections. In Section 2, existing synchronization schemes are reviewed. Proposed reference based tree structured time synchronization scheme is explained in Section 3. Section 4 contains the network evaluation phase, followed by the comparison of proposed scheme with existing work in Section 5. Finally, Section 6 contains the conclusion of the paper.

## 2. EXISTING APPROACHES TO TIME SYNCHRONIZATION

Time synchronization algorithms providing a mechanism to synchronize the local clocks of the nodes in the network have been extensively studied in the past. The most widely adapted protocol used in the internet domain is the NTP devised by Mills [7]. The NTP clients synchronize their clocks to the time servers with accuracy in the order of milliseconds by statistical analysis of the round-trip time. The time servers are synchronized by external time sources, typically using GPS. The NTP has been widely deployed and proved to be effective, secure and robust in the internet. In WSNs, however, non-determinism in transmission time caused by the Media Access Channel (MAC) layer of the radio stack can introduce several hundreds of milliseconds delay at each hop. Therefore, without further adaptation, NTP is suitable only for WSN applications with *low precision* demands.

Two of the most prominent examples of existing time synchronization protocols developed for the WSN domain are the Reference Broadcast Synchronization (RBS) algorithm [9] and the Timing-sync Protocol for Sensor Networks (TPSN) [10].

In RBS, a reference message is broadcasted. The receivers record their local time when receiving the reference broadcast and exchange the recorded times with each other. The main advantage of RBS is that it eliminates transmitter-side non-determinism. The disadvantage of the approach is that additional message exchange is necessary to communicate the local time-stamps between the nodes. In the case of multi hop synchronization, the RBS protocol would lose its accuracy. Santashil PalChaudhuri et al [15] extended the RBS protocol to handle multi hop clock synchronization in which all nodes need not be within single-hop range of a clock synchronization sender.

The TPSN algorithm first creates a spanning tree of the network and then performs pair wise synchronization along the edges. Each node gets synchronized by exchanging two synchronization messages with its reference node one level higher in the hierarchy. The TPSN achieves two times better performance than RBS by time-stamping the radio messages in the MAC layer of the radio stack and by relying on a two-way message exchange. The shortcoming of TPSN is that it does not estimate the clock drift of nodes, which limits its accuracy, and does not handle dynamic topology changes.

TinySeRSync [16] protocol works with the ad hoc deployments of sensor networks. This protocol proposed two asynchronous phases: Phase I –secure single-hop pair wise synchronization, and Phase II–secure andresilient global synchronization to achieve global time synchronization in a sensor network.

Van Greunen et al [18] *Lightweight Tree-based Synchronization* (LTS) protocol is a slight variation of the network-wide synchronization protocol of Ganeriwal et al. [17]. Similar to network-wide synchronization the main goal of the LTS protocol is to achieve reasonable accuracy while using modest computational resources. As with network-wide synchronization, the LTS protocol seeks to build a tree structure within the network. Adjacent tree nodes exchange synchronization information with each other. A disadvantage is that the accuracy of synchronization decreases linearly in the depth of the synchronization tree (i.e., the longest path from the node that initiates synchronization to a leaf node). Authors discuss various ideas for limiting the depth of tree; the performance of protocol is analyzed with simulations.

A survey study of Abolfazl Akbari et al [4] suggested as data communication and various network operations cause energy depletion in sensor nodes and therefore, it is common for sensor nodes to exhaust its energy completely and stop operating. This may cause connectivity and data loss. Therefore, it is necessary that network failures are detected in advance and appropriate measures are taken to sustain network operation. The clock synchronization protocols significantly differ from the conventional protocols in dealing the challenges specific to WSNs. It is quite likely that the choice of a protocol will be driven by the characteristics and requirements of each application.

## 3. TREE STRUCTURED REFERENCING TIME SYNCHRONIZATION APPROACH

In this Section we proposed Tree Structured Referencing Time Synchronization (TSRT) scheme, which is based on the protocol, proposed by [2], that the aim is to minimize the complexity of the synchronization. Thus the needed synchronization accuracy is assumed to be given as a constraint, and the target is to devise a synchronization algorithm with minimal complexity to achieve given precision. TSRT works on two phases. First phase used to construct an ad hoc tree structure and second phase used to synchronize the local clocks of sensor nodes followed by network evaluation phase.

The goal of the TSRT is to achieve a network wide synchronization of the local clocks of the participating nodes. We assume that each node has a local clock exhibiting the typical timing errors of crystals and can communicate over an unreliable but error corrected wireless link to its neighbors. The TSRT synchronizes the time of a sender to possibly multiple receivers utilizing a single radio message time-stamped at both the sender and the receiver sides. MAC layer time-stamping can eliminate many of the errors, as observed in [11]. However, accurate time-synchronization at discrete points in time is a partial solution only. Compensation for the clock drift of the nodes is inevitable to achieve high precision in-between synchronization points and to keep the communication overhead low. Linear regression is used in TSRT to compensate for clock drift as suggested in [9].

### 3.1. Main Ideas

The proposed synchronization approach is flexible and self-organized. A physical broadcast channel is required, which is automatically satisfied by the wireless medium. A connected network is also required in order to spread the synchronization ripple to nodes network wide. The proposed approach assumes the coexistence of reference nodes and normal sensor nodes in a WSN. A *"reference node"* periodically transmits beacon messages to its neighbors. These beacon messages initiate the synchronization waves. Multiple reference nodes are allowed to operate in the system simultaneously. A sensor node in this approach will dynamically select the nearest reference node as its reference for clock synchronization.

This approach exploits the usage of multi-channel radios to improve precision, minimize the communication overhead and low energy consumption. A common control channel is shared by all the sensor nodes for delivery of beacon messages and control packets. This control channel can be the same one as is used for general data traffic. Each sensor node is also assigned a unique clock channel different from all its neighbors' clock channels. Usage of a dedicated clock channel reduces the variation in propagation delay caused by packet collisions and retransmissions, thereby improving the accuracy of clock estimation.

This proposed protocol used for multi-hop synchronization of the network based on pair wise synchronization scheme suggested by [10]. This requires nodes to synchronize to some *reference point(s)* such as a sink node in the sensor network and needs a tree to be constructed first. Then pair wise synchronization is done along the *n - 1* edges of the tree. In such algorithms, the reference node is the root of the tree and has the responsibility of initiating a "resynchronization" when required. Using the assumption that the clock drifts are bounded, and given the required precision, the reference node calculates the time period that a single synchronization step will be valid. Since the depth of the tree affects the time to synchronize the whole network, and also the precision error at the leaf nodes, the depth of the tree is communicated back to the root node so that it can use this information in its resynchronization time decision.

An explanation of a standard two-way message exchange between a pair of nodes [13] employing for Synchronization is helpful to understand proposed synchronization design. The basic building block of the synchronization process is the two-way message exchange between a pair of nodes Here we assume that the clock drift between a pair of nodes is constant in the small time period during a single message exchange. The propagation delay is also assumed to be constant in both directions. Consider a two-way message exchange between nodes A and B as shown in Fig. 1. Node A initiates the synchronization by sending a synchronization message at time $t_1$ as per node's local clock. This Message includes A's identity, and the value of $t_1$. B receives this message at $t_2$ which can be calculated as

$$t_2 = t_1 + \Delta + d \qquad (1)$$

Where $\Delta$ is the relative clock drift between the nodes, and d is the propagation delay of the pulse.

B responds at time $t_3$ with an acknowledgement, which includes the identity of B and the values

$t_1$, $t_2$, and $t_3$. Then, node A can calculate the clock drift and propagation delay as below, and synchronize itself with respect to node B.

$$\Delta = ((t_2 - t_1) - (t_4 - t_3))/2 \qquad (2)$$

$$d = ((t_2 - t_1) + (t_4 - t_3))/2 \qquad (3)$$

The synchronization phase is initiated by the root node's syn_begin message. On receiving this message, nodes of level 1 initiate a two-way message exchange with the root node. Before initiating the message exchange, each node waits for some random time, in order to minimize collisions on the wireless channel. Once they get back a reply from the root node, they adjust their clocks to the root node. Level 2 nodes, overhearing some level 1 node's communication with the root, initiate a two-way message exchange with a level 1 node, again after waiting for some random time to ensure that level 1 nodes have completed their synchronization. This procedure eventually gets all nodes synchronized with reference to the root node, the synchronization process described in detail in Subsection 3.3.

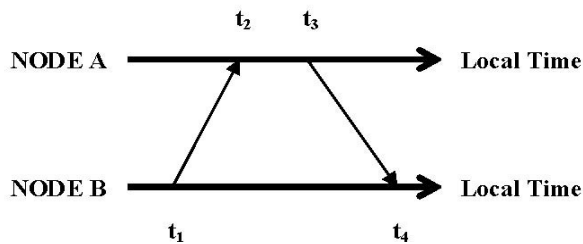

Figure 1. Two way message exchange between a pair of nodes.

### 3.2. Ad Hoc Tree Construction Phase:

Before the sensors can be synchronized, a tree structure based network topology must be created. The proposed Algorithm 1 is used by each sensor node to efficiently flood the network to form a logical hierarchical structure from a designated source point. Each sensor is initially set to accept fd_pckt (flood packets) for first time, but will ignore subsequent ones in order not to be continuously reassigned as the flood broadcast propagates. When a node receives or accepts the fd_pckt then first it set to its parent as source of broadcast after that level of current receiver node will be assigned one more than the level of parent node and then it broadcast the fd_pckt along with node identifier and level. If a node receives the ack_pckt, the variable *no_receiver* increments to keep track of the node's receivers.

Algorithm 1. Tree Structure Construction

```
Begin
   Accept (fd_pckts)
   Initialize : no_reciever = 0;
   Node_Level(Root)=0;
   If (current_reciever = = root)
       Broadcast (fd_pckts)
    Else if (current_reciever != root)
       Begin
            Accept (fd_pckts);
            Parent(curent_reciever) = Source(broadcast_msg);
            Node_Level(curent_reciever)=Node_Level(Parent)+1;
            Broadcast (ack_pckt, node_id);
```

```
              Ignore (fd_pckts);
        End
          Else if (current_node receives ack_pckt)
              no_ receiver++;
  End
```

### 3.3. Hierarchical Time Synchronization Phase

The first component of TSRT's bidirectional time synchronization service is the push-based [12] Hierarchy Time Synchronization (HTS) Scheme. The goal of HTS is to enable central authorities to synchronize the vast majority of a WSN in a lightweight manner. This approach particularly based on pair wise synchronization with allusion to single reference node is discussed in Subsection 3.1.

#### 3.3.1. Single Reference Node

As shown in Fig. 2, HTS consists of three simple steps that are repeated at each level in the hierarchy. First, a Reference Node (RN) broadcasts a beacon on the control channel (Fig. 2A). One child node specified by the reference node will jump to the specified clock channel, and will send a reply on the clock channel (Fig. 2B). The RN will then calculate the clock offset and broadcast it to all child nodes, synchronizing the first ripple of child nodes around the reference node (Fig. 2C). This process can be repeated at subsequent levels in the hierarchy further from the reference node (Fig. 2D). The HTS scheme is explained in more detail as follows:

```
Step 1: RN initiates the synchronization by broadcasting
the syn_begin message with time t1 using the control
channel and then jumps to the clock channel. All concerned
nodes   record   the   received   time   of   the   message
announcement. RN randomly specifies one of its children,
e.g. SN2, in the announcement. The node SN2 jumps to the
specified clock channel.
Step 2: At time t3, SN2 replies to the RN with its
received times t2 and t3.
Step 3.1: RN now contains all time stamps from t1 to t4.
It calculates clock drift ∆ and propagation delay d, as
per equation (2) and (3), and calculate t2 = t1 + ∆ + d,
and then broadcasts it on the control channel.
Step 3.2: All involved neighbor nodes, (SN2, SN3, SN4 and
SN5) compare the time t2 with their received timestamp
t2'.
 i.e. SN3 calculates the offset d' as:
       d' = t2 – t2'
Finally, the time on SN3 is corrected as:
       T = t + d +d'
Where t is the local clock reading.
Step 4: SN2, SN3, SN4 and SN5 initiate the syn_begin to
their downstream nodes.
```

We assume that each sensor node knows about its neighbors when it initiates the synchronization process. In Step 1, the response node is specified in the announcement. It's the only node that jumps to the clock channel specified by the RN. The other nodes are not disturbed by the synchronization conversation between RN and SN2 and can conduct normal data communication while waiting for the second update from the RN. A timer is set in the RN when the syn_begin message is transmitted. In case the message is lost on its way to SN2, the RN goes back to the normal control channel after the timer expires and thus avoids indefinite waiting.

As the synchronization ripple spreads from the reference node to the rest of the network, a multi-level hierarchy is dynamically constructed. Levels are assigned to each node based on its distance to reference node, i.e. number of hops to the central reference point. Initiated from the reference nodes, the synchronization steps described above are repeated by the nodes on each level from the root to the leaves.

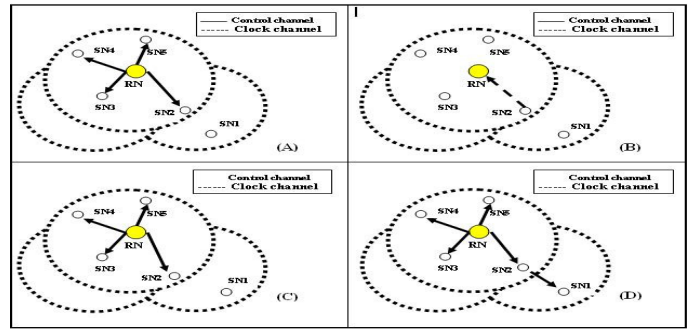

Figure 2. (A) Reference node broadcasts (B) A neighbor replies (C) All neighbors are synchronized (D) Repeat at lower layers

## 4. NETWORK EVALUATION PHASE

In this phase, the network examines the total amount of message exchanges for synchronization during the last synchronization period, and then adjusts the duration of the next synchronization period to minimize the overall energy consumption for synchronization. As per [14] when the network traffic occurs rarely and synchronization delay is not a critical problem, applying the sensor initiated (SI) mode is a better choice to save network resources instead of using the always on (AO) mode. In addition, for some applications, the sensor clocks might be allowed to go out of synchronization unless sensing events happen. Another critical problem is to determine the required number of timing message exchanges (beacons) per pair wise synchronization. To fulfill higher requirement of synchronization accuracy, a larger number of message transfers and corresponding signal processing is needed for pair wise synchronization. However, as the number of required timing messages per pair wise synchronization increases, the overall number of timing messages in a synchronization period increases. Hence, there is a tradeoff between accuracy and energy consumption.

To address these design challenges, we consider various factors to deter-mine network parameters such as the synchronization mode, the re-sync period $\tau$, and the number of beacons per pair wise synchronization $N$. Indeed, it is aiming at efficient usage of network resources (message exchanges) in synchronization. The network parameters are summarized as follows:

$B$: number of branches in a Ad hoc tree of the network.
$\tau$: re-sync period
$h$: average number of hops per unit time
$\delta$: latency factor reflecting the amount of allowed delay in data transmission
$N$: number of beacons per pair wise synchronization

The number of branches in the network $B$ can be obtained after the Ad hoc Tree Construction phase. The latency factor $\delta$ should be fixed according to the type of a sensor network and its range is from 0 to 1. The higher latency factor means higher concern for network delays. In every sensing event, its destination node adds the number of hops that have occurred in that particular transmission to its storage. During the synchronization phase, the reference node collects the information of the total number of hops occurred in the last synchronization period and determines the average number of hops per unit time ($h$) in the network. This information can be included in timing messages with a small overhead. As mentioned, the goal of this phase is to minimize the number of required timing messages. In [14] suggested that the number of timing messages per unit time is given by $M=2BN/\tau$ in AO mode, while in the SI mode, $M=2hN$. To minimize the number of timing messages per unit time $M$, the synchronization mode should be selected as follows:

$$2BN\delta/\tau <> 2hN \qquad (4)$$

Where the latency factor $\delta$ varies from 0 to 1 such that more delay dependant networks assumes a larger value of $\delta$ and vise versa ($0 \leq \delta \leq 1$). For example, $\delta$ is set to be 0 for sensor networks requiring network synchronization all the time. On the other extreme, for delay-independent networks, $\delta$ should be close to 1. As the clock synchronization period $\tau$ increases, the network becomes more power efficient. Thus, $\tau$ should be chosen as large as possible. However, a too large value of $\tau$ induces a critical synchronization problem since the clock difference (offset) between nodes keeps generally increasing with time. Hence, there exists a maximum timing synchronization period ($\tau_{max}$) which is determined by the oscillator regulations (hardware specifications) and the accuracy of estimators. Notice that sensing data transmission is not available during the synchronization phase ($\tau_{sync}$), so the re-sync period $\tau = \tau_{max} + \tau_{sync}$. In continuation, (4) can be rewritten as

$$\tau < B\delta/h \qquad (5)$$

From (5), the synchronization mode changes from AO into SI when $\tau$ is smaller than $B\delta/h$ and vise versa. In the SI mode, the reference node periodically asks the number of hops that occurred during the past time interval, and then make a decision whether or not to switch to the AO mode. Actually, $\tau$ is also dependent on $N$ since it strongly depends on the accuracy of timing offset estimators. A more detailed analysis of $\tau$ is provided in next Subsection.

### 4.1. Optimum Number of Beacons (N) and Resynchronization Period ($\tau$)

The number of timing messages (beacons) per pair wise synchronization ($N$) is a critical parameter to determine both the synchronization accuracy and power efficiency. Suppose that the clock timing mismatch $\varepsilon$ between the two nodes is modeled as follows: $\varepsilon = \varepsilon_o + \varepsilon_s t$, where $t$ denotes the reference time, $\varepsilon_o$ and $\varepsilon_s$ stand for the clock offset and skew errors, respectively. Let $\varepsilon_{o,i}$ and $\varepsilon_{s,i}$ denote the clock offset and skew estimation errors when $i$ message exchanges occur between the two nodes. In general, it is difficult to determine any specific mathematical model for either clock offset or skew errors. In this paper, we model both clock offset and skew errors by normal distributions based on the experimental results reported in [9], [10]:

$$\varepsilon_{o,i} = N(0, \sigma^2_{\varepsilon o,i}) \quad 1 \leq i \leq N,$$

$$\varepsilon_{s,i} = N(0, \sigma^2_{\varepsilon s,i}) \quad 1 \leq i \leq N,$$

where $\varepsilon_{s,1}$ stands for the clock skew error when no skew estimation occurred. Then, the maximum clock mismatch can be modeled another normal distribution $\varepsilon = N(0, \sigma^2_\varepsilon)$, where $\sigma^2_\varepsilon = \sigma^2_{\varepsilon o,n} + \sigma^2_{\varepsilon s,n} \tau_{max}$, ($t = \tau_{max}$). Imposing the upper-limit $\varepsilon_{max}$ for the clock error via the probabilistic measure:

$$Ps = Pr(|\varepsilon| \leq \varepsilon_{max}) = \text{erfc}(\varepsilon_{max}/\sqrt{2}\sigma_\varepsilon)$$

where $\text{erfc}(x) \equiv (2/\sqrt{\pi}) \int_x^\infty e^{-t^2} dt$ and $Ps$ denotes the network wide sync error probability. Thus, $\sigma_\varepsilon$ can be determined when $\varepsilon_{max}$ and the limit of sync error probability are fixed. For instance,

27

when *Ps* is limited to 0.1% and $\varepsilon_{max}$ is 10 *ms*, $\sigma_\varepsilon$ becomes 3.04 *m*.
The maximum timing sync period with *N* beacons can be written as

$$\tau_{max} = \sqrt{((\sigma^2_\varepsilon - \sigma^2_{\varepsilon_{o,N}})/ \sigma^2_{\varepsilon_{s,i,N}})} \qquad (6)$$

Based on the lower bounds and asymptotic performance of the estimators, one can easily infer closed-form expressions of the variances $\varepsilon_{o,N}$ and $\varepsilon_{s,N}$ in terms of the variances $\varepsilon_{o,1}$ and $\varepsilon_{s,2}$ respectively.

## 5. ANALYSIS OF TSRT

The TSRT protocol exploits the broadcast nature of the wireless medium to establish a single common point of reference, i.e. the arrival time of the broadcast message is the same on all neighbor peers. This common reference point can be used to achieve synchronization in Step 4 of the Subsection 3.3, i.e. t2 at node SN2 occurred at the same instant as t2' at node SN3. As the RN is synchronizing itself with SN2, the other neighboring nodes can overhear the RN's initial broadcast to SN2 as well as the RN's final update informing SN2 of its offset d2. If in addition the RN includes the time t2 in the update sent to SN2 (redundant for SN2), then that allows all neighbors to synchronize.

The intuition is that, since t2 and t2' occurred at the same instant, then overhearing t2 gives SN3 its offset from SN2's local clock and overhearing d2 gives SN3 the offset from SN2's local clock to the RN reference clock. Thus, SN3 and all children of the RN can calculate their own offsets to the RN reference clock with only three messages (2 control broadcasts and 1 clock channel reply) TSRT is highly scalable and lightweight, since there is only one lightweight overhead exchange per hop between a parent node and all of its children. In contrast, synchronization in RBS happens between a pair of neighbors, which is called pair verification, rather than between a central node and all of its neighbors.

As a result, RBS is susceptible to high overhead as the number of peers increases [1]. The TSRT approach eliminates the potential broadcast storm that arises from pair wise verification, while at the same time preserving the advantage of reference broadcasting, namely the common reference point. Also, since the TSRT parent provides the reference broadcast that is heard by all children, then TSRT avoids the problem in RBS when two neighbors of an initiating peer are "hidden" from each other. The parameters used in the protocol dynamically assign the hierarchy level to each node during the spread of the synchronization ripple and no additional routing protocol is required. TSRT is lightweight since the number of required broadcasting messages is constant in one broadcasting domain.

Only three broadcast messages are necessary for one broadcasting domain, no matter how dense the sensor nodes are. In TSRT, the sender error is eliminated by comparing the received time on each node.

TSRT' current policy for selecting the child node to respond to the sync begin message is a random selection. However, it is possible to incorporate historical knowledge from previous TSRT cycles in the selection of the next child responder. Moreover, previous TSRT responses may be combined to broadcast a composite value in Step 3. This may be useful to account for propagation delay differences between neighbors within a local broadcast domain, which we can assumed to be small, but which may become more relevant when the distances between neighbors becomes very large in a highly distributed WSN.

The performance comparison of TSRT and TPSN in terms of the average number of message exchanges M with respect to the number of beacons N is shown in fig. 3 and fig. 4 when Ps is assumed as 0.01% and 1% respectively. This simulation is based on the linear network model where the depth of the network *B* = 5, $\varepsilon_{max}$ =10*ms*, $\sigma_{\varepsilon_o}$ = 16.67μ, *d* = 10*ms*, *t* = 400*ms*, and $\sigma_{\varepsilon_s}$

=1.58μ have assumed.

It can be observed that TSRT requires a less number of timing messages than TPSN when there multiple numbers of beacon transmissions are required. Moreover, the gap of the average number of required timing messages between TSRT and TPSN significantly increases as N increases, and thus TSTP is by far more efficient than TPSN for large value of N. It can be also seen that a few number of beacons is enough to minimize M for TSRT. Besides, as expected, a larger number of beacons required to meet a more strict constraint of the network-wide error probability Ps. In practice, a lower number of N is highly preferable since it is proportional to the synchronization time, i.e., a lower N induces better latency performance. Although, it may not be optimal in terms of energy consumption.

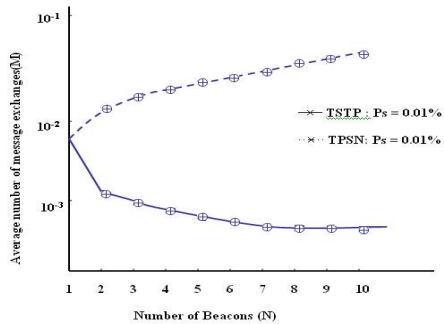
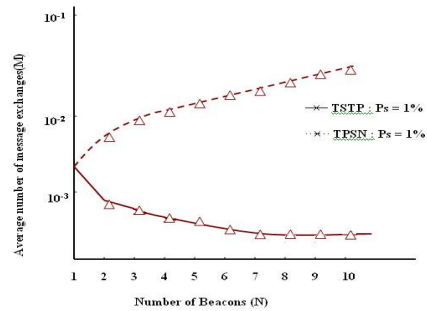

Figure 3.  Average Number of Messages (M) for Ps=0.01%

Figure 4.  Average Number of Messages (M) for Ps=1%

## 6. CONCLUSION

WSN have tremendous advantages for monitoring object movement and environmental properties but require some degree of synchronization to achieve the best results. The proposed TSRT synchronization approach is able to produce deterministic synchronization with only few pair wise message exchanges.

While the proposed approach is especially useful in WSN which are typically, extremely constrained on the available computational power, bandwidth and have some of the most exotic needs for high precision synchronization. The proposed approach was designed to switch between TPSN and RBS. These two algorithms allow all the sensors in a network to synchronize themselves within a few microseconds of each other, while at the same time using the least amount of resources possible. In this work two varieties of the algorithm are presented and their performance is verified theoretically with the existing results and compared with existing protocols.  The comparison with RBS and TPSN shows that the proposed synchronization approach is lightweight since the number of required broadcasting messages is constant in one broadcasting domain.

### ACKNOWLEDGEMENTS

The authors are thankful to all synchronization papers that are credited as references below. This paper contains comprehensive study of papers on synchronization algorithms or approaches collectively, cited from various contributions from the authors mentioned below.

**Authors**

*Surendra Rahamatkar* has received his Bachelor of Engineering in Computer Science & Engineering from Barkatullah University, Bhopal, India, Post Graduate Diploma in Advanced Computing from CDAC, Pune and Master in Technology in Computer Science & Engineering from VMRF Deemed University, India. He is a member of various Technical Societies viz. CSI, International Association of Engineers (IEA), ISTE. He is internationally recognized as a Member of Editorial Board of International Journal of Computer Applications (IJCA), New York, USA, International Journal of Advanced Engineering & Applications and Reviewer of International Journal of Computer Theory and Engineering (IJCTE), Singapore & ICMLC 2011 Singapore. He published many research papers in various International/ National Journals and Conferences. His main research interests include: Wireless Sensor Network, Distributed & Mobile Computing and Middleware.

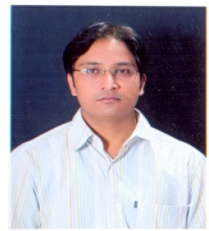

*Dr. Ajay Agarwal* has done B.Tech. Degree in Computer Science & Engineering from Institute of Engineering & Technology, Lucknow (India), M.Tech.(honors) Degree in Computer Science & Engineering from Motilal Nehru Regional Engineering College, Allahabad and Ph.D. in Computer Science from Indian Institute of Technology, Delhi (India). Presently he is working as a Professor and Head in department of Computer Applications at Krishna Institute of Engineering & Technology, Ghaziabad, India. He is a member of various Technical Societies viz. Institute of Electrical and Electronics Engineers (IEEE), Computer Society of India (CSI), Indian Society for Technical Education (ISTE), Institution of Engineers India, Institute of Chartered Computer Professional of India and Indian Association of Physics Teachers. He published many papers in various International/ National Journals and Conferences. His main research interests include: Wireless Sensor Network, Mobile Computing and Middleware.

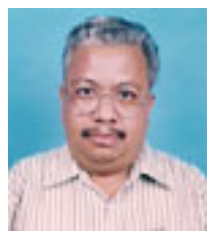